\documentclass[aps,prd,preprint,preprintnumbers,nofootinbib]{revtex4-1}
\usepackage{graphicx}
\usepackage[utf8]{inputenc} 
\usepackage{amsmath}
\usepackage{amssymb}
\usepackage{float}
\usepackage{comment}
\usepackage{soul}
\usepackage[usenames,dvipsnames]{color}

\newcommand{\UAN}{Centro de Investigaciones, Universidad Antonio Nari\~{n}o,\\
Carrera 3 Este \# 47A-15, Bogot\'{a}, Colombia}
\newcommand{\UPTC}{Escuela de Física, Universidad Pedagógica y Tecnológica de Colombia,\\
Avenida Central del Norte \# 39-115, Tunja, Colombia}
\newcommand{\UdeA}{Instituto de Física, Universidad de Antioquia,\\Calle 70 \# 52-21, Apartado Aéreo 1226, Medellín, Colombia}
\newcommand{\mzprime}{M_{Z'}}
\newcommand{\ICTP}{Simons Associate at The Abdus Salam International Centre for Theoretical Physics (ICTP),\\
  Strada Costiera 11, 34151, Trieste, Italy}

\usepackage{slashed} 
\usepackage{hyperref}

\begin{document}
\title{Two-component dark matter and a massless neutrino \\in a new $\boldsymbol{B-L}$ model}

\author{Nicolás Bernal~${}^1$} \email{nicolas.bernal@uan.edu.co} 
\author{Diego Restrepo~${}^{2,\,3}$}\email{restrepo@udea.edu.co}
\author{Carlos Yaguna~${}^4$}\email{carlos.yaguna@uptc.edu.co}
\author{Óscar Zapata~${}^2$}\email{oalberto.zapata@udea.edu.co}

\affiliation{$^1$~\UAN}
\affiliation{$^2$~\UdeA}
\affiliation{$^3$~\ICTP}
\affiliation{$^4$~\UPTC}

\begin{abstract}
We propose a new extension of the Standard Model by a  $U(1)_{B-L}$ gauge symmetry in which the anomalies are canceled by two right-handed neutrinos  plus four chiral fermions with fractional $B-L$ charges. Two scalar fields that break the $B-L$ symmetry and give masses to the new fermions are also required. After symmetry breaking, two neutrinos acquire Majorana masses via the seesaw mechanism leaving a massless neutrino in the spectrum. Additionally, the other new fermions arrange themselves into two Dirac particles, both of which are automatically stable and contribute to the observed dark matter density. This model thus realizes in a natural way, without ad hoc discrete symmetries, a two-component dark matter scenario. We analyze in some detail the dark matter phenomenology of this model. The dependence of the relic densities with the parameters of the model is illustrated and the regions consistent with the observed dark matter abundance are identified. Finally, we impose the current limits from LHC and direct detection experiments, and show that the high mass region of this model remains unconstrained.

  \end{abstract}

\begin{flushright}
        {\tt PI/UAN-2018-634FT}
\end{flushright}

\maketitle

\section{Introduction}
One of the main problems in particle physics today is to find out what is the New Physics that lies beyond the Standard Model (SM). For a long time, supersymmetric models were considered the most promising candidates; they are well-motivated theoretically and, thanks to the plethora of new supersymmetric particles, 
they give rise to multiple experimental signals that may be observed in current detectors. So far, however, none of these signals has actually been detected. The LHC, in particular, has not found any evidence of supersymmetric particles (see e.g. Refs.~\cite{Sirunyan:2017cwe,Aaboud:2017dmy,Ventura:2017itv}), casting  doubt on their existence. Nowadays, non-supersymmetric models seem to be preferred as candidates for New Physics.

Among them, those that can account for neutrino masses and dark matter (DM) are clearly favored.  Oscillation experiments have established, beyond reasonable doubt, the existence of non-zero neutrino masses, a fact that cannot be explained within the SM~\cite{deSalas:2018bym}. Cosmological observations, on the other hand, indicate the existence of an exotic form of matter, dubbed DM, that is not made up of any known particle~\cite{Aghanim:2018eyx}. Given that the evidence for neutrino masses and DM requires New Physics beyond the SM, it makes sense to focus our attention in models that can simultaneously solve both of these problems. In addition, it would be helpful if this New Physics also gives rise to observable signals in current experiments, including the LHC.  

Models based on an extra gauge symmetry of $B-L$, $U(1)_{B-L}$, fit the bill. They all include a new gauge boson ($Z'$) that couples to both leptons and quarks, inducing detectable signals at colliders such as the LHC~\cite{Basso:2008iv}. They naturally lead to a realization of the seesaw mechanism of neutrino mass generation~\cite{Mohapatra:1979ia}, which is the most appealing way of explaining the smallness of neutrino masses. They can also easily accommodate scalar or fermion DM~\cite{Okada:2010wd,Okada:2012sg,Sanchez-Vega:2015qva,Rodejohann:2015lca,Patra:2016ofq,Klasen:2016qux}. Additionally, they are well-motivated theoretically as they often appear in GUT theories based on the $SO(10)$ group~\cite{Fritzsch:1974nn}.

Different realizations of the  $U(1)_{B-L}$ extension have been considered in the literature~\cite{Montero:2007cd,Okada:2010wd,Guo:2015lxa,Patra:2016ofq} but they all require additional fermions charged under $B-L$ to cancel the anomalies. The most common   way to do so is to include three right-handed neutrinos (with $B-L$ charge equal to $-1$), which usually play also a role in neutrino mass generation. Models without right-handed neutrinos have also been studied~\cite{Sanchez-Vega:2015qva,Ma:2015mjd,Patra:2016ofq}. In Ref.~\cite{Patra:2016ofq}, for instance, the anomalies are canceled by four chiral fermions with fractional $B-L$ charges that help explain the DM. 

In this paper we put forward a novel realization of the $U(1)_{B-L}$ extension in which the anomalies are canceled partially by two right-handed neutrinos and partially by the DM particles. The crucial point is that current neutrino data requires only the existence of two massive neutrinos, which can be achieved via the seesaw mechanism, with just two right-handed neutrinos rather than the usual three. To cancel entirely the anomalies we then add four chiral fermions with fractional $B-L$ charges. After symmetry breaking, these fermions arrange themselves into two Dirac particles, both of which turn out to be automatically stable and to contribute to the DM density. This model, therefore, realizes a two-component DM scenario (see e.g. Refs.~\cite{Zurek:2008qg,Profumo:2009tb,Esch:2014jpa,Arcadi:2016kmk, Ahmed:2017dbb}) in a natural way, without any discrete symmetry.  In the scalar sector,  the model contains only two additional scalar fields that break the $B-L$ symmetry and give masses to the new fermions. We study the phenomenology of this model in some detail, with particular emphasis on the DM aspects.  Current bounds from colliders and DM direct detection experiments are also analyzed.

The rest of the paper is organized as follows. In section~\ref{sec:model} the model is introduced and the free parameters are identified. The generation of SM neutrino masses is discussed in section~\ref{sec:neutrino}. The dependence of the relic densities with the parameters of the model is illustrated in section~\ref{sec:pheno}. In section \ref{sec:viable} we show the results of an intensive scan over the parameter space of the model. The viable regions are characterized and the constraints from collider and direct detection experiments are imposed. Finally, we draw our conclusions in section~\ref{sec:conc}.

\section{The model}
\label{sec:model}
 
 We propose a model that extends  the gauge symmetry of the SM with an additional $U(1)$ of baryon minus lepton number ($B-L$), that is based on  $SU(3)_c\times SU(2)_L\times U(1)_Y\times U(1)_{B-L}$. With just the SM fermions, this model is not anomaly-free as both the anomaly with three $B-L$ gauge bosons and the gravitational anomaly  with one $B-L$ gauge boson turn out to be different from zero. The usual way of canceling them is by adding three right-handed neutrinos, which are  singlets of the SM and have charge $-1$ under $B-L$. Here, we suggest instead a novel way of canceling the anomalies with  six fields: two right-handed neutrinos plus four chiral fermions, which are singlets of the SM and have  fractional $B-L$ charges.  These fractional charges are not unique. In our model we take them to be $10/7$, $-4/7$, $-2/7$ and $-9/7$ respectively for the fields $\xi_L$, $\eta_R$, $\zeta_R$ and $\chi_L$, where $L$ and $R$ denote the chirality.  It is straightforward to check that the anomalies indeed cancel with this assignment for the six chiral fields.
 
Regarding the scalar sector, the model contains only two new fields ($\phi_{1,\,2}$), both singlets  of the SM and with $B-L$ charges $1$ and $2$, respectively. These two scalars are enough to spontaneously break the $B-L$ symmetry and to give masses to all the new fermions.  The full particle content of our model with their respective charges is displayed in Table~\ref{tab:particles}. 
\begin{table}
  \centering
  \begin{tabular}{|c||c|c|}
\hline
Particles & $U(1)_{B-L}$ & $\left(SU(3)_c,\,SU(2)_L,\,U(1)_Y \right)$ \\\hline\hline
$Q_{Li}$ & $1/3$  & $ \left(\mathbf{3},\,\mathbf{2},\,1/6  \right)$  \\
$u_{Ri}$ & $1/3$  & $ \left(\overline{\mathbf{3}},\,\mathbf{1},\,2/3  \right)$  \\
$d_{Ri}$ & $1/3$  & $ \left(\overline{\mathbf{3}},\,\mathbf{1},\,-1/3  \right)$  \\ \hline
 $L_i$ & $-1$ & $ \left(\mathbf{1},\,\mathbf{2},\,-1/2\right)$  \\
$e_{Ri}$ & $-1$  & $ \left(\mathbf{1},\,\mathbf{1},\,-1  \right)$  \\ 
 $N_{R1}$ & $-1$ & $ \left(\mathbf{1},\,\mathbf{1},\,0\right)$  \\ 
 $N_{R2}$ & $-1$ & $ \left(\mathbf{1},\,\mathbf{1},\,0\right)$  \\ \hline
 $\xi_{L}$    & $10/7$ & $ \left(\mathbf{1},\,\mathbf{1},\,0\right)$ \\
 $ \eta_{R} $ & $-4/7$ & $ \left(\mathbf{1},\,\mathbf{1},\,0\right)$ \\
  $\zeta_{R}$    & $-2/7$ & $ \left(\mathbf{1},\,\mathbf{1},\,0\right)$ \\
 $\chi_{L} $ & $-9/7$ & $ \left(\mathbf{1},\,\mathbf{1},\,0\right)$ \\\hline
 $H$ & 0  & $ \left(\mathbf{1},\,\mathbf{2},\,1/2\right)$\\ 
 $\phi_1$ &  $1$   & $ \left(\mathbf{1},\,\mathbf{1},\,0\right)$ \\
 $\phi_2$ &  $2$   & $ \left(\mathbf{1},\,\mathbf{1},\,0\right)$ \\
\hline
  \end{tabular}
  \caption{The full particle content of our model with their quantum numbers. The index $i$ denotes the SM fermion generations and goes from $1$ to $3$.}
  \label{tab:particles}
\end{table}

The most general Lagrangian involving  the new fields and consistent with the $SU(3)_c\times SU(2)_L\times U(1)_Y\times U(1)_{B-L}$ gauge symmetry  contains the following terms
\begin{align}
\label{eq:TheModel}
\mathcal{L}&\supset \sum_{F}i\,\overline{F}\left(\slashed{\partial} +  i\,g_{BL}\,q_{F}\,Z_\mu^\prime \gamma^\mu \right)F - \left(a\, \overline{\xi_{L}}\, \eta_{R}\,\phi_2 + b\, \overline{\zeta_{R}}\, \chi_{L}\, \phi_1+ \text{h.c.} \right)+\left(\frac{y_1}{2}\bar N_{R1}N_{R1}^c\phi_2^\dagger \right. 
\nonumber \\ &\,+ \left.\frac{y_2}{2}\bar N_{R2}N_{R2}^c\phi_2^\dagger +\text{h.c.}\right)+\sum_{S}\left|\left( \partial_\mu +i\,g_{BL}\,q_S\,Z'_\mu \right) S\right|^2- \mathcal{V}-\left(y_{ij}\overline{L}_{Li}\tilde{H}\nu_{Rj}+ \text{h.c.} \right),
\end{align}
where $g_{BL}$ is the gauge coupling associated to the $U(1)_{B-L}$ group and $Z_\mu^\prime$ is its corresponding gauge boson. $F$ and $S$ denote the new chiral fermions and new scalars respectively, and $q_{F,\,S}$ their $B-L$ charges. $a$, $b$, $y_1$ and $y_2$ are new Yukawa couplings involving the new fields whereas $y_{ij}$ ($i=1$, 2, 3 but $j=1$, 2) are the usual Yukawa couplings between right-handed neutrinos, the lepton doublets and the SM Higgs boson.

The new  scalar  potential reads 
 \begin{align}
\mathcal{V}&= \mu^2_H  H^\dagger H + \lambda_H (H^\dagger H)^2 
     + \mu^2_1 \phi^\dagger_1 \phi_1 + \lambda_1 (\phi^\dagger_1 \phi_1)^2 
      +\mu^2_2 \phi^\dagger_2 \phi_2 + \lambda_2 (\phi^\dagger_2 \phi_2)^2  \nonumber \\
      &+\rho_{1} (H^\dagger H) (\phi^\dagger_1 \phi_1)
      +\rho_{2} (H^\dagger H) (\phi^\dagger_2 \phi_2)
      +\lambda_{3} (\phi^\dagger_1 \phi_1) (\phi^\dagger_2 \phi_2) 
      +\mu \left( \phi_2 \phi^{\dagger^2}_1 + \phi_2^\dagger \phi^{^2}_1 \right).
      \label{eq:V}
\end{align} 
The conditions for this potential to be bounded from below are
\begin{equation}
\label{eq:vacuum}
 \lambda_H,\,\lambda_1,\,\lambda_2 \geq 0,\;\;
 \rho_1 + \sqrt{\lambda_H \lambda_1} \geq 0\, , \;\; \rho_2 + \sqrt{\lambda_H \lambda_2} \geq 0\,\;\;\text{ and }\;\;\lambda_3 + \sqrt{\lambda_1 \lambda_2} \geq 0 \,.
\end{equation}

The spontaneous symmetry breaking of $SU(2)_L \times U(1)_Y \times U(1)_{B-L}$ down to $SU(2)_L \times U(1)_Y$ is achieved by assigning non-zero vacuum expectation values (vevs) to the scalars $\phi_1$ and $\phi_2$ at a scale above the  electroweak phase transition scale. Later, $SU(2)_L \times U(1)_Y$ breaks down to electromagnetism via the neutral component of the Higgs doublet, $H$. 

The fields $H^0$, $\phi_1$ and $\phi_2$ can be parameterized in terms of real scalars and pseudoscalars as
\begin{align}
&H^0 =\frac{1}{\sqrt{2} }(v+h)+  \frac{i}{\sqrt{2} } G^0\, , \nonumber \\
& \phi_1 = \frac{1}{\sqrt{2} }(v_1+h_1)+  \frac{i}{\sqrt{2} } A_1\,, \nonumber \\
& \phi_2 = \frac{1}{\sqrt{2} }(v_2+h_2)+  \frac{i}{\sqrt{2} } A_2\, ,
\end{align}
with $\langle H^0\rangle=v/\sqrt2$, $\langle \phi_1\rangle=v_1/\sqrt2$ and $\langle \phi_2\rangle=v_2/\sqrt2$. The minimization conditions of the scalar potential imply that
\begin{align}
\mu^2_H& = -\left(\lambda^2_H v^2 + \frac{ \rho_1}{2} v^2_1 +   \frac{ \rho_2}{2} v^2_2 \right),  \nonumber \\
\mu^2_1& = -\left(\lambda^2_1 v^2_1 + \frac{ \rho_1}{2} v^2 +   \frac{ \lambda_3}{2} v^2_2+\sqrt{2} v_2 \mu \right) , \nonumber \\
\mu^2_2& = -\left(\lambda^2_2 v^2_2 + \frac{ \rho_2}{2} v^2 +   \frac{ \lambda_3}{2} v^2_1 +\frac{1}{\sqrt{2}} \frac{v^2_1 \mu}{v_2} \right).  \nonumber 
\end{align}

Because  $\phi_1$ and $\phi_2$ are charged under $B-L$, their vevs induce a non-zero mass for the neutral gauge boson $Z^\prime$ associated with the $B-L$ gauge symmetry. This mass is given by  
\begin{eqnarray}
	M^2_{Z^\prime}= g^2_{BL} \left( v_1^2+ 4 v_2^2 \right).
\end{eqnarray}
It is convenient to define a new dimensionless parameter, $\tan\beta$, as the ratio between the vevs of the scalars fields $\phi_1$ and $\phi_2$: $\tan\beta\equiv\frac{v_1}{v_2}$. Thus, 
\begin{equation}
M^2_{Z^\prime}= g^2_{BL} v_2^2 \left(4+\tan^2\beta\right),
\end{equation}
so that $v_1$ and $v_2$ can be written in terms of $M_{Z'}$, $g_{BL}$ and $\tan\beta$.

Since the $Z^\prime$ couples to the SM fermions, its mass and coupling can be constrained with collider data. From LEP II the bound reads~\cite{Carena:2004xs,Cacciapaglia:2006pk} 
\begin{equation}
\label{eq:LEPII}
\frac{M_{Z'}}{g_{BL}}\gtrsim 7~\mathrm{TeV}.
\end{equation}

Going back to the scalar potential, Eq.~\eqref{eq:V}, notice that the terms proportional to $\rho_1$ and $\rho_2$ induce mixing between the SM Higgs boson and the two new scalar fields. Since the scalar boson observed at the LHC with a mass of $M_h\simeq 126$~GeV is very much SM-like, this mixing is necessarily small. For simplicity, in the following we will neglect it, effectively setting $\rho_{1,\,2}$ to zero.  

The scalar CP-even spectrum thus consist of the SM Higgs plus two other states which mix with each other  according to the mass matrix
\begin{align}
\label{eq:scalarMass}
\mathcal{M}^2_{\mbox{\small CP-even}} =
\begin{pmatrix} 
2 \lambda_1 v^2_1     &   v_1( \lambda_3  v_2 + \sqrt{2}  \mu )   \\    
v_1( \lambda_3  v_2 + \sqrt{2} \mu )   &   2 \lambda_2 v^2_2 -\frac{\mu\,v^2_1}{\sqrt{2}v_2}
\end{pmatrix}
\end{align}
in the $(h_1,\,h_2)$ basis. The resulting mass eigenstates, denoted by $H_1$ and $H_2$, are related to $h_{1,\,2}$ via the mixing angle, $\theta$:
\begin{equation}
\begin{pmatrix} h_1\\h_2\end{pmatrix}=\begin{pmatrix}\cos\theta &\sin\theta\\
-\sin\theta &\cos\theta\end{pmatrix}\begin{pmatrix}H_1\\H_2\end{pmatrix}.
\end{equation}
It is convenient to take as free parameters of the scalar sector the physical masses of $H_{1,\,2}$ ($M_{H_{1,\,2}}$) and the mixing angle $\theta$.  The couplings $\lambda_i$ can then be expressed in terms of them as
\begin{align}
\lambda_1=& \frac{1}{2v_1^2}\left[\cos^2\theta\,M^2_{H_1}+\sin^2\theta\,M^2_{H_2}\right],\\
\lambda_2=& \frac{1}{2v_2^2}\left[\sin^2\theta\,M^2_{H_1}+\cos^2\theta\,M^2_{H_2}+\frac{\mu\,v_1^2}{\sqrt 2v_2}\right],\\
\lambda_3  =&\frac{1}{v_1v_2}\left[\sin\theta\cos\theta\,(M_{H_2}^2-M_{H_1}^2)-\sqrt 2\mu\,v_1\right].
\end{align}

The mass matrix for the CP-odd scalars in the basis ($A_1$, $A_2$) is given instead by
\begin{align}
\mathcal{M}^2_{\mbox{\small CP-odd}} = 
\begin{pmatrix} 
-2 \sqrt{2} v_2\,\mu                &  \sqrt{2} v_1\,\mu      \\
\sqrt{2} v_1\,\mu                   &   -\frac{v^2_1}{\sqrt{2}} \frac{\mu}{v_2}  
\end{pmatrix},
\end{align}
and, as expected, has an eigenvalue equal to zero --the would-be Goldstone boson that becomes the longitudinal mode of the $Z^\prime$. The mixing angle, $\alpha$, in this sector is defined by 
\begin{equation}
\begin{pmatrix} A_1\\A_2\end{pmatrix}=\begin{pmatrix}\cos\alpha &\sin\alpha\\
-\sin\alpha &\cos\alpha\end{pmatrix}\begin{pmatrix}G'\\A\end{pmatrix},
\end{equation}
where $G'$ is the Goldstone whereas $A$ is the physical CP-odd scalar.  The mixing angle $\alpha$ is entirely determined by the vevs according to  $\sin\alpha= -\sqrt{4v_2^2/(v_1^2+4v_2^2)}$. It is convenient to take the mass of  $A$, $M_A$, as a free parameter of the model. The parameter $\mu$ is then expressed as
\begin{equation}
\mu = -\frac{M_A^2\,\sin^2\alpha}{2\sqrt 2 \,v_2}\,.
\end{equation}

This model predicts, therefore, the existence of 3  scalar fields beyond the SM Higgs: $H_1$, $H_2$ and $A$. These fields have scalar interactions among themselves, gauge interactions with the $Z^\prime$, and Yukawa interactions with the new fermions.

These new fermions all become massive after the spontaneous breaking of $B-L$. The two right-handed neutrinos acquire Majorana masses (denoted by $M_{R1}$ and $M_{R2}$) from the Lagrangian terms proportional to $y_{1,\,2}$ whereas the remaining four chiral fermions form two Dirac particles, which we will denote by $\psi_1$ and $\psi_2$:  
\begin{align}
  \psi_1=\xi_{L}+\eta_{R}\,\,\,\,\text{and}\,\,\,\, \psi_2=\chi_{L}+\zeta_{R}.
\end{align}
Their masses are given by
\begin{align}
  M_1=a\frac{v_2}{\sqrt 2},\,\,\,\, M_2=b\frac{v_1}{\sqrt 2},
\label{eq:dmmasses}
\end{align}
respectively for $\psi_1$ and $\psi_2$. 

From the Lagrangian one can see that $\psi_1$ and $\psi_2$ are both automatically stable. In fact, the model is \emph{accidentally} invariant under two independent $\mathbb{Z}_2$ symmetries: one under which $\xi_L$ and $\eta_R$ are odd while the rest are even; and another under which the odd particles are instead $\chi_L$ and $\zeta_R$. That is the reason why we get two stable particles. Besides being stable, $\psi_1$  and $\psi_2$ are neutral under the SM gauge group, which renders them  viable DM candidates. This model realizes, therefore, a two-component DM scenario.

All in all, this model introduces $17$ additional parameters: four masses for the new fermions ($M_{R1}$, $M_{R2}$, $M_1$, $M_2$), three scalar masses ($M_{H1}$, $M_{H2}$, $M_A$), one mixing angle in the scalar sector ($\sin\theta$), six neutrino Yukawa couplings ($y_{ij}$), the ratio of the two vevs ($\tan\beta$), the gauge coupling constant  ($g_{BL}$) of the $U(1)_{B-L}$ group, and the mass of the new gauge boson $M_{Z'}$.  These parameters are constrained by a combination of collider, neutrino, and DM experiments. 

\section{Neutrino masses}\label{sec:neutrino}
In this model neutrino masses are generated via a seesaw mechanism with two right-handed neutrinos~\cite{Ibarra:2003up}. As a consequence,  a massless active neutrino should be present in the spectrum.  If all three neutrinos were found to be massive this model would be excluded.

The Majorana masses of the two  right-handed neutrinos appear after the breaking of the $B-L$ symmetry and are given by
\begin{align}
M_{R1}=y_1\frac{v_2}{\sqrt 2}\,,\qquad M_{R2}=y_2\frac{v_2}{\sqrt 2}\,,
\end{align}
 and are therefore expected to be below $v_2$. $v_2$ contributes to the mass of the $Z'$ and induces the mass of $\psi_1$, one of the two DM particles present in this model. As we will see in the next sections, the DM constraint requires these masses (and consequently $v_2$) to be  around the TeV scale. Thus, we actually have a TeV scale realization of the seesaw mechanism~\cite{Han:2006ip,delAguila:2007qnc,Atre:2009rg,Ibarra:2011xn} with two right-handed neutrinos.
 
 The light neutrino mass matrix is given by the usual seesaw formula,
 \begin{align}
 m_\nu=-\frac{v^2}{2}\,y\,M_R^{-1}\,y^T,
 \end{align}
 which can be inverted with the Casas-Ibarra parameterization~\cite{Casas:2001sr} to express the Yukawa couplings, $y_{ij}$, in terms of measurable quantities (neutrino masses, mixing angles and phases) and one additional complex angle~\cite{Ibarra:2003up}. This scenario is thus, by construction, consistent with current neutrino data.

\section{Dark matter phenomenology}
\label{sec:pheno}

\begin{figure}[t]
\begin{center}
\begin{tabular}{ccc}
  \includegraphics[scale=0.4]{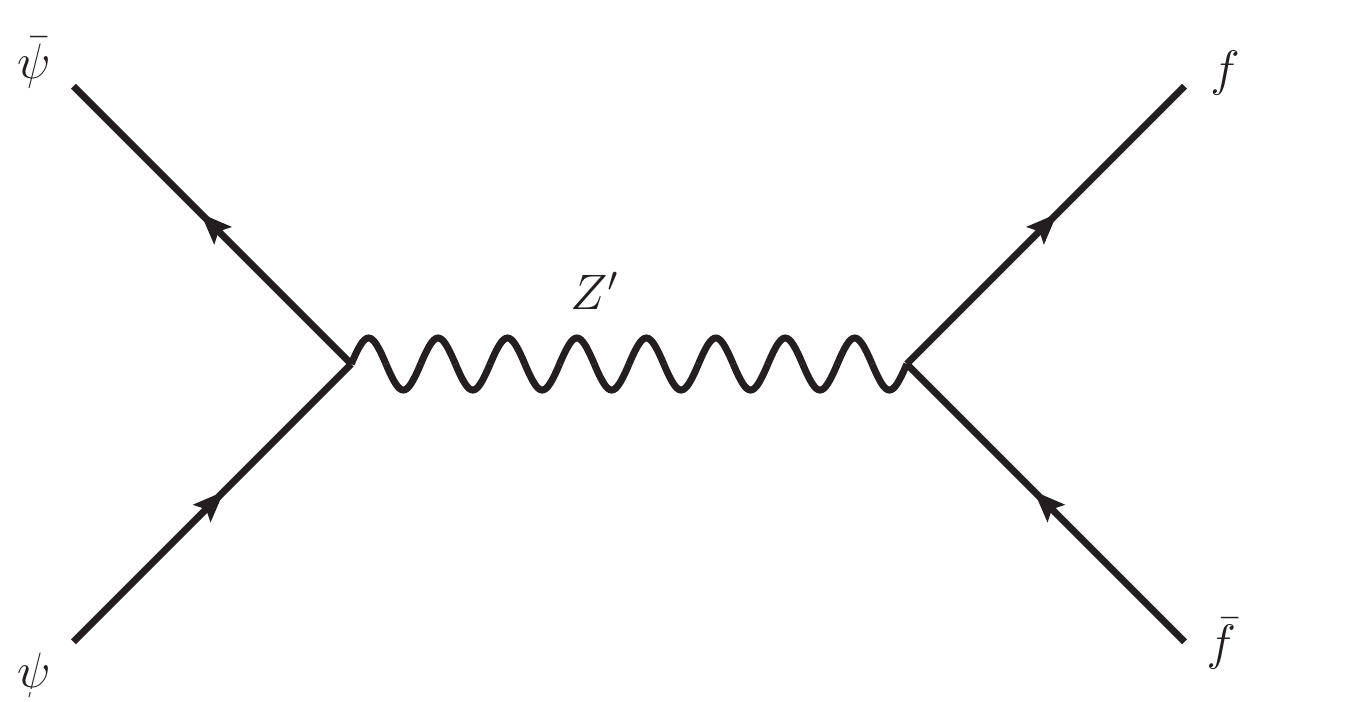} &
  \includegraphics[scale=0.4]{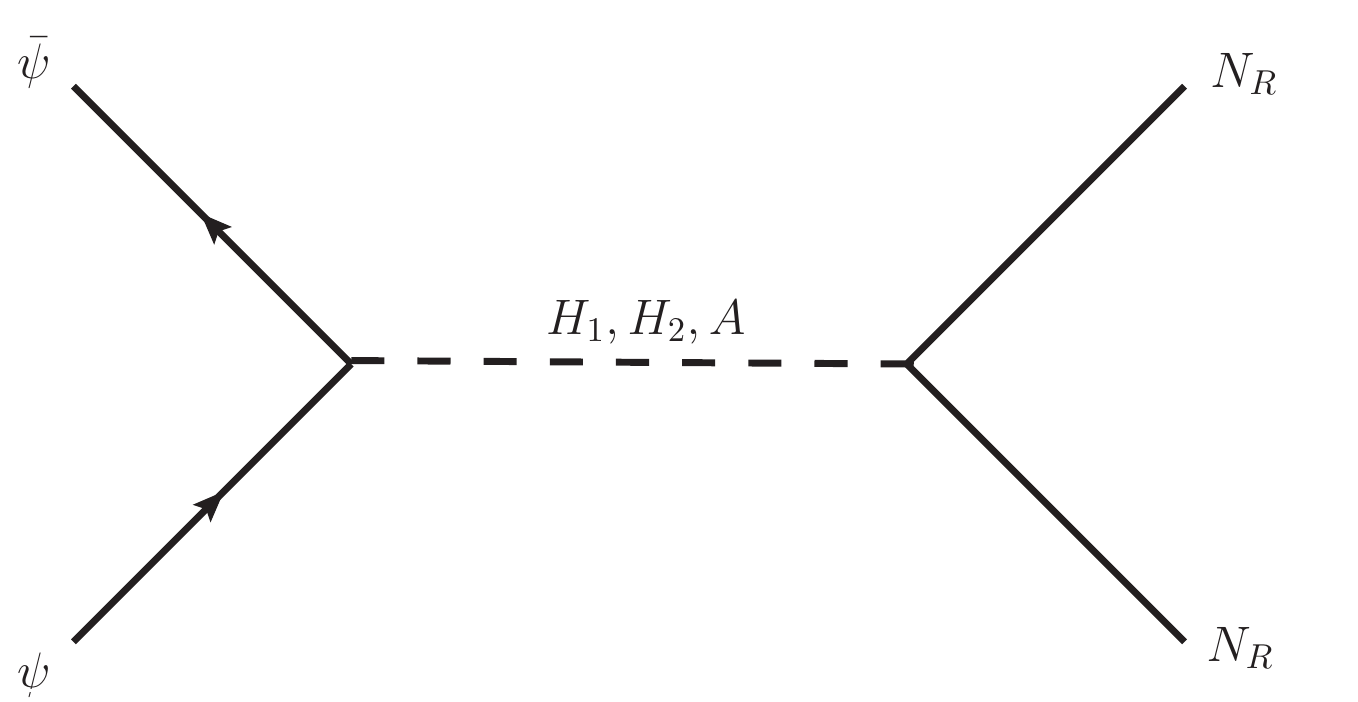} &
  \includegraphics[scale=0.4]{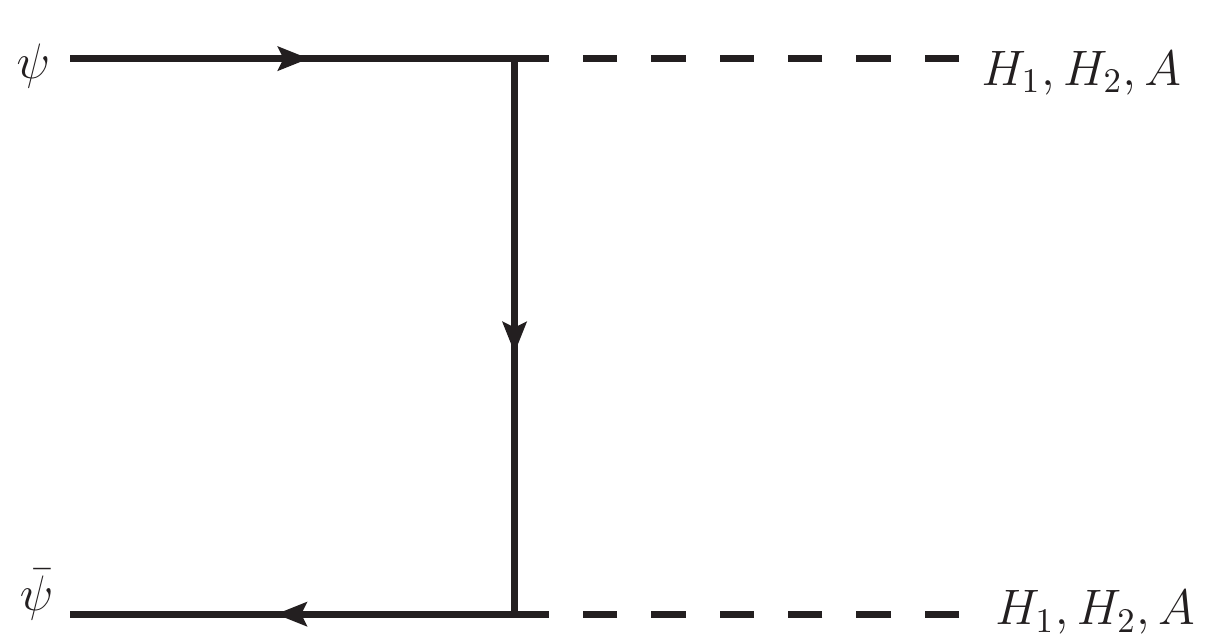}\\
  (a) & (b) & (c)
\end{tabular}
\caption{Some of the Feynman diagrams that contribute to DM annihilation in this model.}
\label{fig:feynman}
\end{center}
\end{figure}

A remarkable feature of this model is that it automatically incorporates two DM particles,  $\psi_1$ and $\psi_2$, that both contribute to the DM density.  Their relic abundances are denoted respectively as $\Omega_1$ and $\Omega_2$ and  their sum should coincide with the  observed DM density, $\Omega_\text{DM}h^2\equiv\Omega_1h^2+\Omega_2h^2\simeq 0.12$~\cite{Aghanim:2018eyx}. $\psi_1$ and $\psi_2$ are Dirac fermions and they  interact with the  new scalars and with the $Z'$. The vector and axial couplings between the DM particles and the $Z'$ can be read off the Lagrangian and are given by
\begin{align}
\label{eq:dmcouplings}
g_{\psi_1V}&= -\frac 37\,g_{BL}, \quad g_{\psi_1A}=g_{BL},\\
g_{\psi_2V}&= \frac{11}{14}\,g_{BL},\quad g_{\psi_2A}=-\frac{g_{BL}}{2}.
\end{align}
They can be used to understand semi-quantitatively different  DM observables. The  annihilation of the DM particles into SM fermions mediated by the $B-L$ gauge boson --Fig.~\ref{fig:feynman} (a)--, for instance, has a cross section that, in the non-relativistic limit and neglecting fermion masses, is given by   
\begin{align}
\sigma v \left(\overline{\psi_i} \psi_i \to Z^{\prime^*} \to \bar{f}f \right)
&=
\frac{N_c^f\,g^2_{fV}\,g^2_{\psi_i V}\,M^2_i 
}{\pi
\big[ \left(4 M^2_i- M^2_{Z^\prime} \right)^2 + M^2_{Z^\prime} \Gamma^2_{Z^\prime}\big] 
 } \,,
 \label{eq:cs}
\end{align}
where $\Gamma_{Z'}$ is the total decay width of the $Z'$ whereas $(N_c^f,\,g_{fV})$ is equal to $(1,\,-g_{BL})$ for leptons and to $(3,\,g_{BL}/3)$ for
quarks. Thus, for $M_1=M_2$, and provided that the scalar interactions can be neglected, the relic densities of the two DM particles will be related by $\Omega_2/\Omega_1\approx g_{\psi_1V}^2/ g_{\psi_2V}^2\approx 0.3$. That  is, $\psi_1$ and $\psi_2$ would account, respectively, for about $75\%$ and $25\%$ of the observed DM density. In this case, the annihilation final states are  determined by the $B-L$ quantum numbers and are given, in order of increasing importance, by charged leptons, neutrinos, right-handed neutrinos (if kinematically allowed), and quarks, with the same contribution from each flavor.

The DM particles in this model can also annihilate into right-handed neutrinos via the $s$-channel exchange of scalar mediators --see Fig.~\ref{fig:feynman} (b). Notice that in such a diagram all of the new particles of this model play a role. The new fermions appear as initial and final states whereas the new scalars are the mediators. These scalars can also appear as final states, Fig.~\ref{fig:feynman} (c), a process that, as will be seen,  is more relevant at high DM masses. Other possible final states include $Z'Z'$ and $SZ'$, where $S=H_1$, $H_2$ and $A$.

To accurately compute the DM relic density, including all possible final states, we have implemented this model into LanHEP \cite{Semenov:2008jy}  and MicrOMEGAs \cite{Belanger:2014vza}, which since its version 4.1 has the capability of dealing with two DM particles.   As a check, we also  implemented the model, independently, in SARAH~\cite{Staub:2011dp, Staub:2013tta, Staub:2015kfa} and verified that our results were consistent. 

\begin{figure}[t]
\begin{center}
  \includegraphics[scale=0.65]{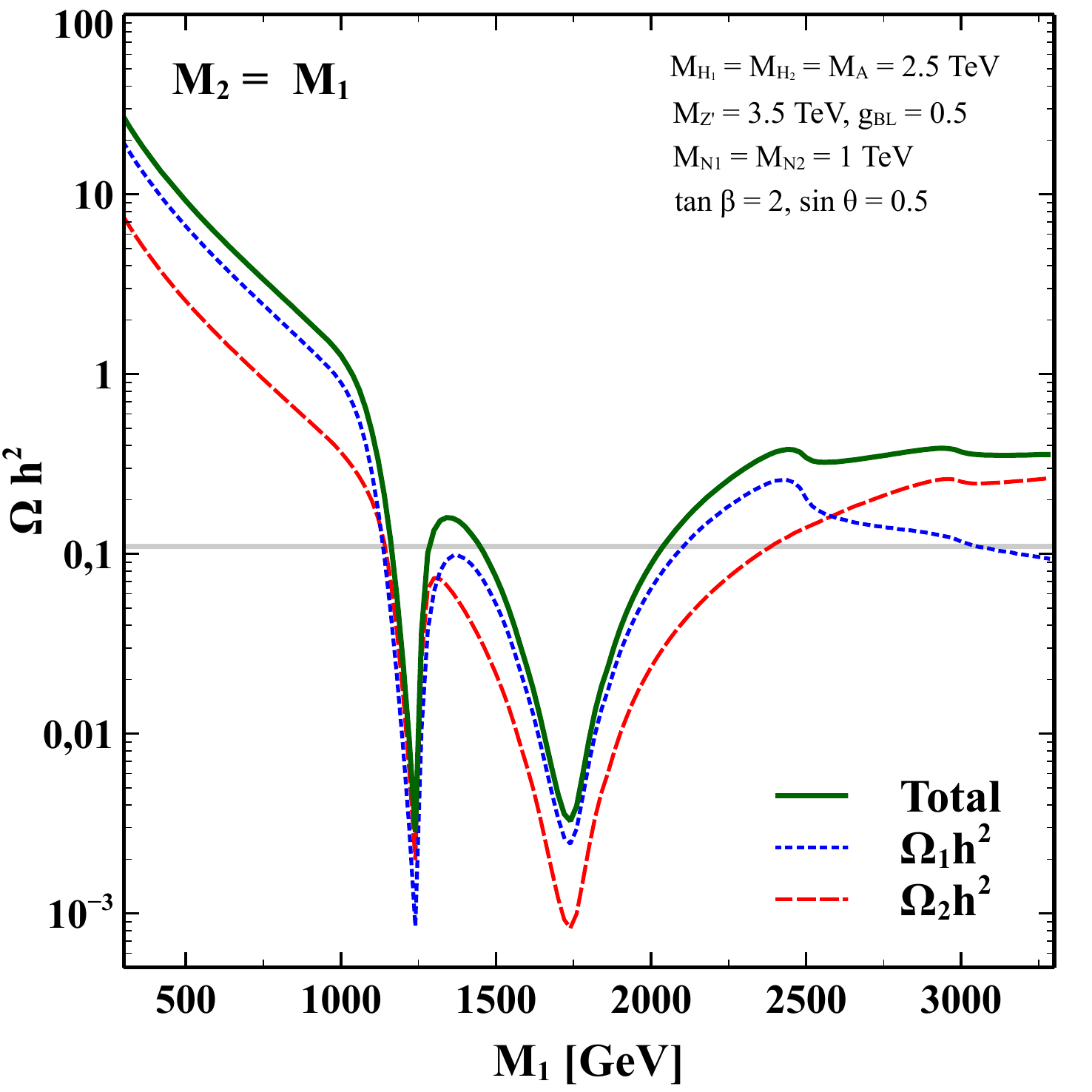}
\caption{\small The relic densities as a function of the $\psi_1$ mass for $M_1=M_2$. The other parameters were taken to be $M_{H_1}=M_{H_2}=M_A= 2.5$ TeV, $M_{N_1}=M_{N_2}=1$ TeV, $\mzprime=3.5$ TeV, $g_{BL}=0.5$, $\tan\beta =2$ and $\sin\theta = 0.5$.  The solid (green) line shows the total relic density, $(\Omega_1+\Omega_2)h^2$, while the dotted (blue) and dashed (red) lines show respectively $\Omega_1h^2$ and $\Omega_2h^2$. The horizontal band is the region consistent with Planck data.}
\label{fig:omega3}
\end{center}
\end{figure}

\begin{figure}[t]
\begin{tabular}{lr}
  \includegraphics[scale=0.5]{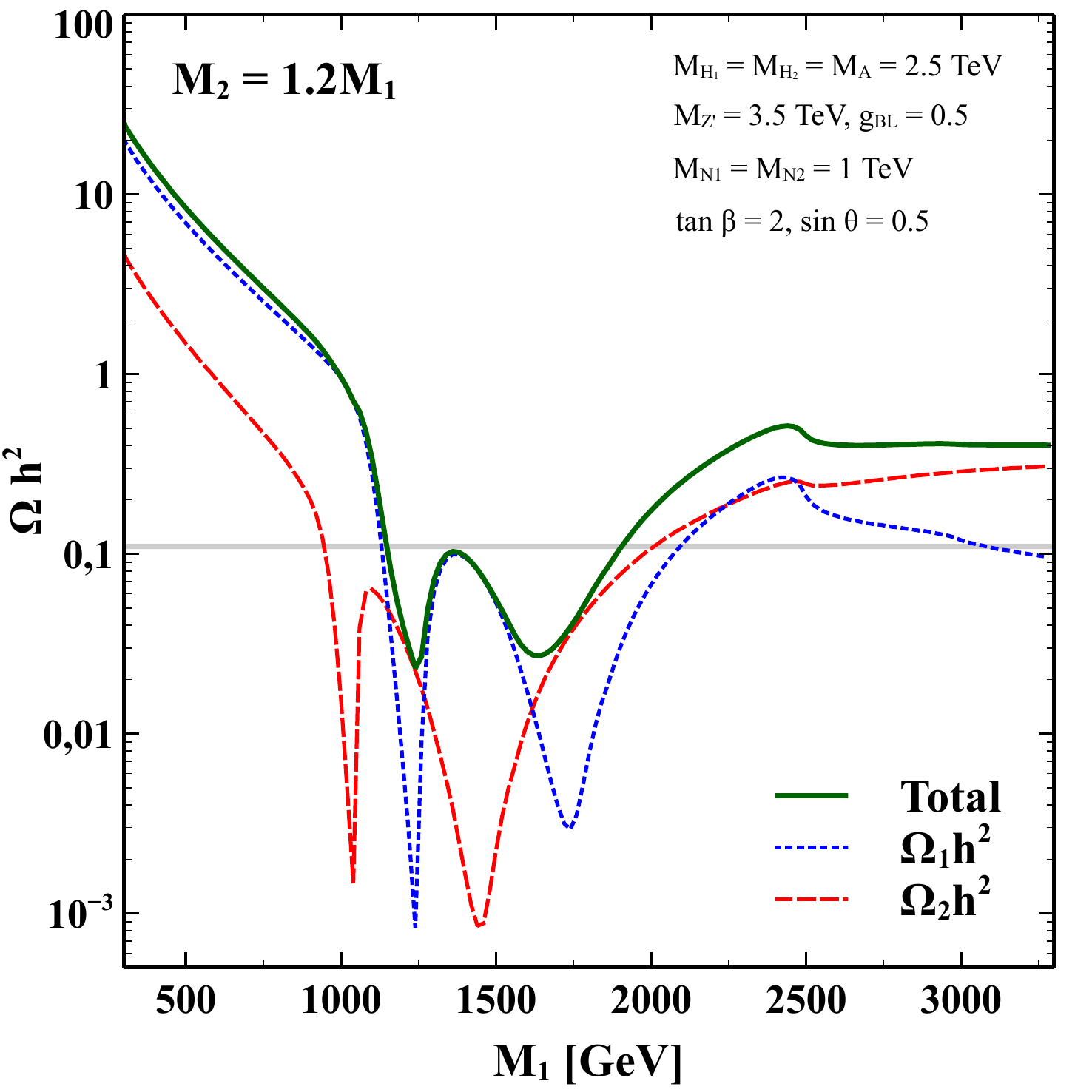} & \includegraphics[scale=0.5]{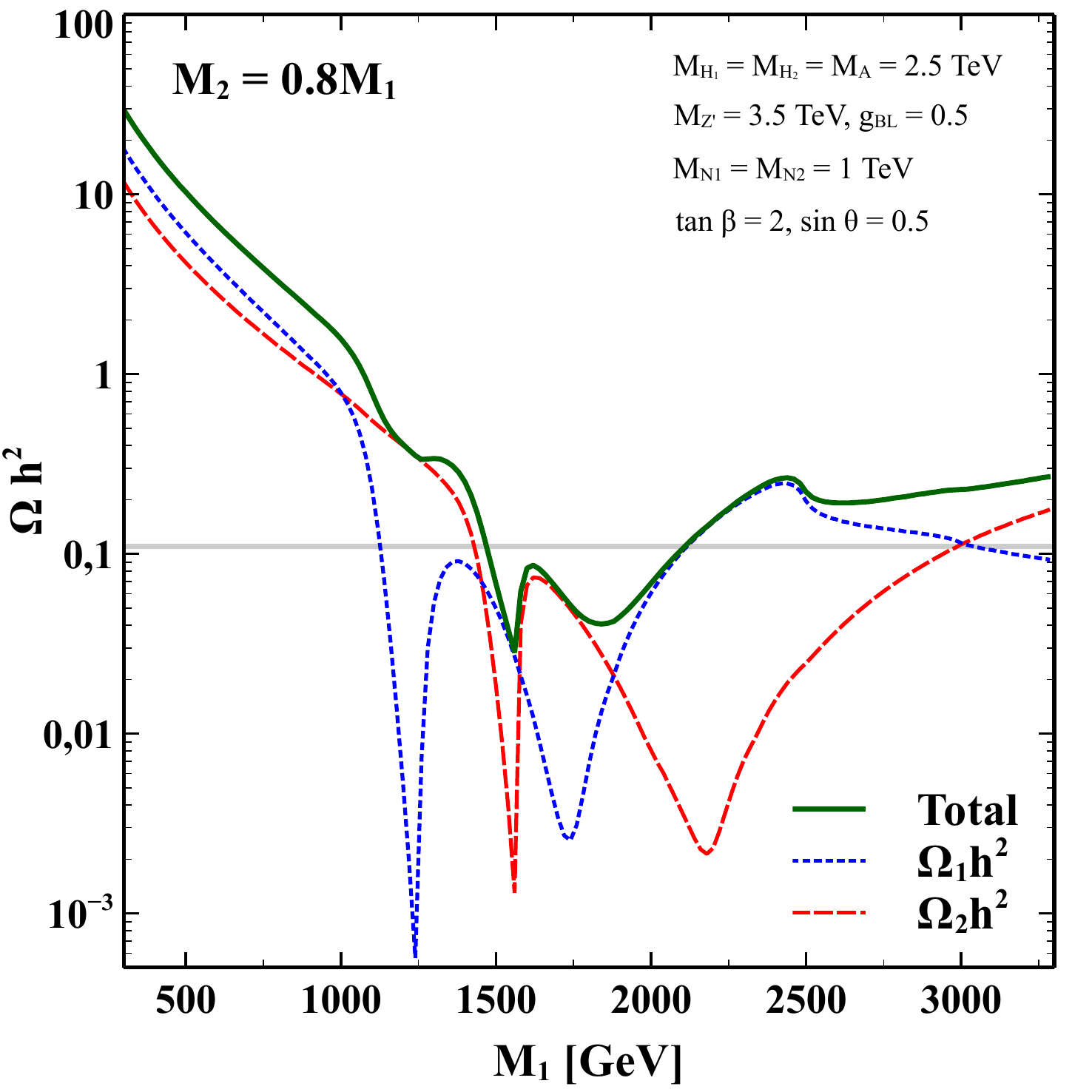}
\end{tabular}
\caption{\small The relic densities as a function of the $\psi_1$ mass for $M_2=1.2\,M_1$ (left) and $M_2=0.8\,M_1$ (right). The other parameters were taken to be $M_{H_1}=M_{H_2}=M_A= 2.5$~TeV, $M_{N_1}=M_{N_2}=1$~TeV, $\mzprime=3.5$~TeV, $g_{BL}=0.5$, $\tan\beta =2$ and $\sin\theta = 0.5$.  The solid (green) line shows the total relic density, $(\Omega_1+\Omega_2)h^2$, while the dotted (blue) and dashed (red) lines show respectively $\Omega_1h^2$ and $\Omega_2h^2$. The horizontal band is the region consistent with Planck data.}
\label{fig:omega1}
\end{figure}

\begin{figure}[h!]
\begin{tabular}{lcr}
  \includegraphics[scale=0.37]{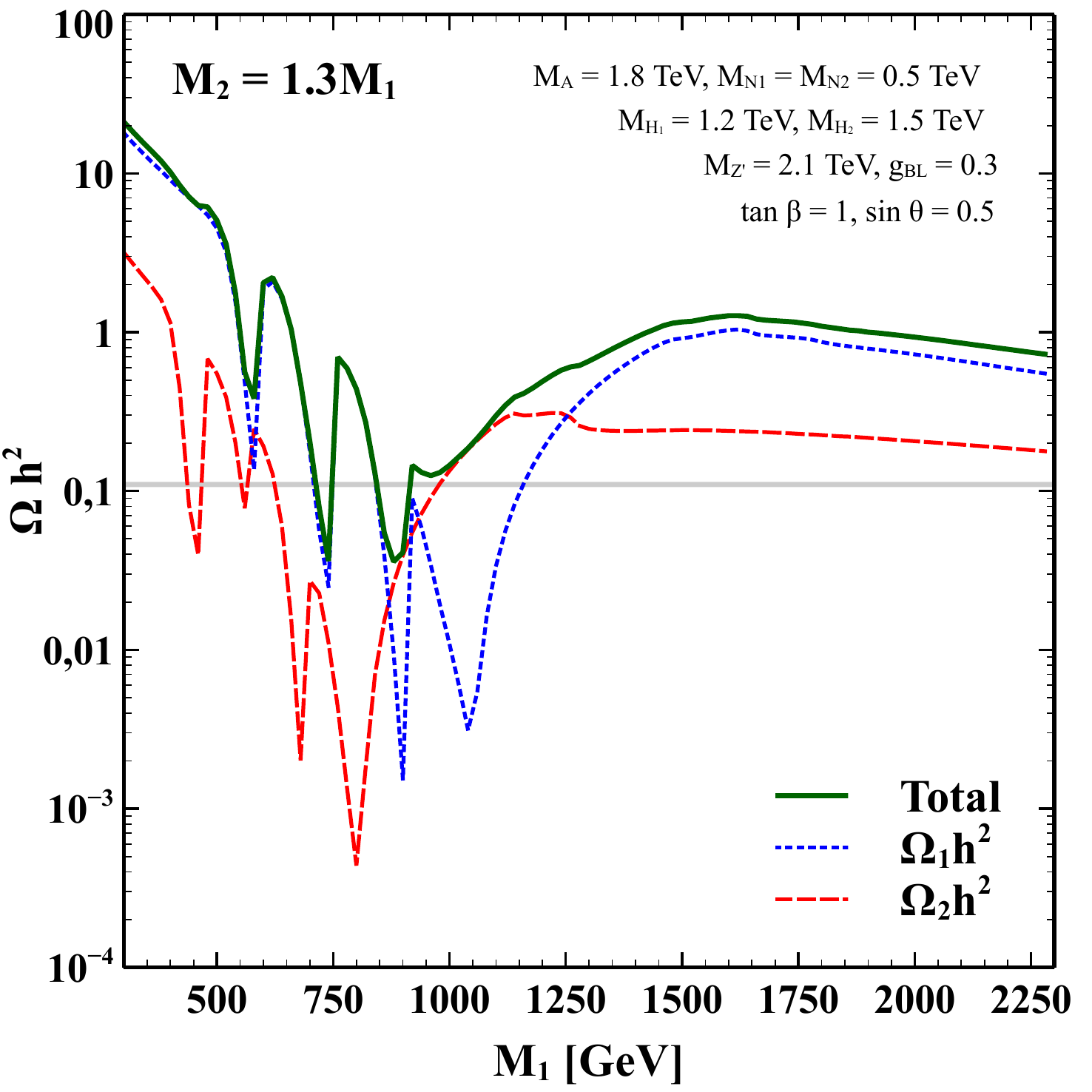} & \includegraphics[scale=0.37]{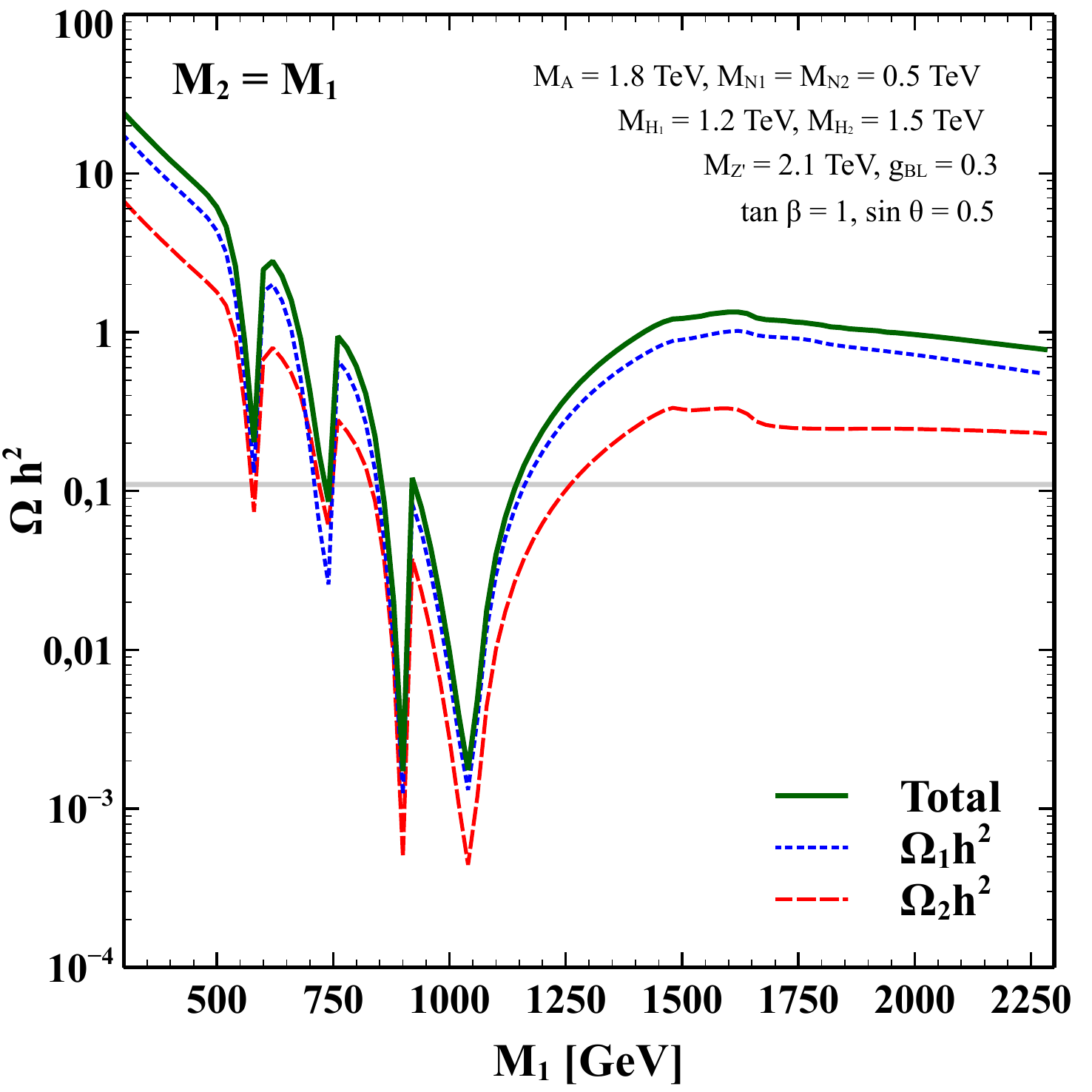}& \includegraphics[scale=0.37]{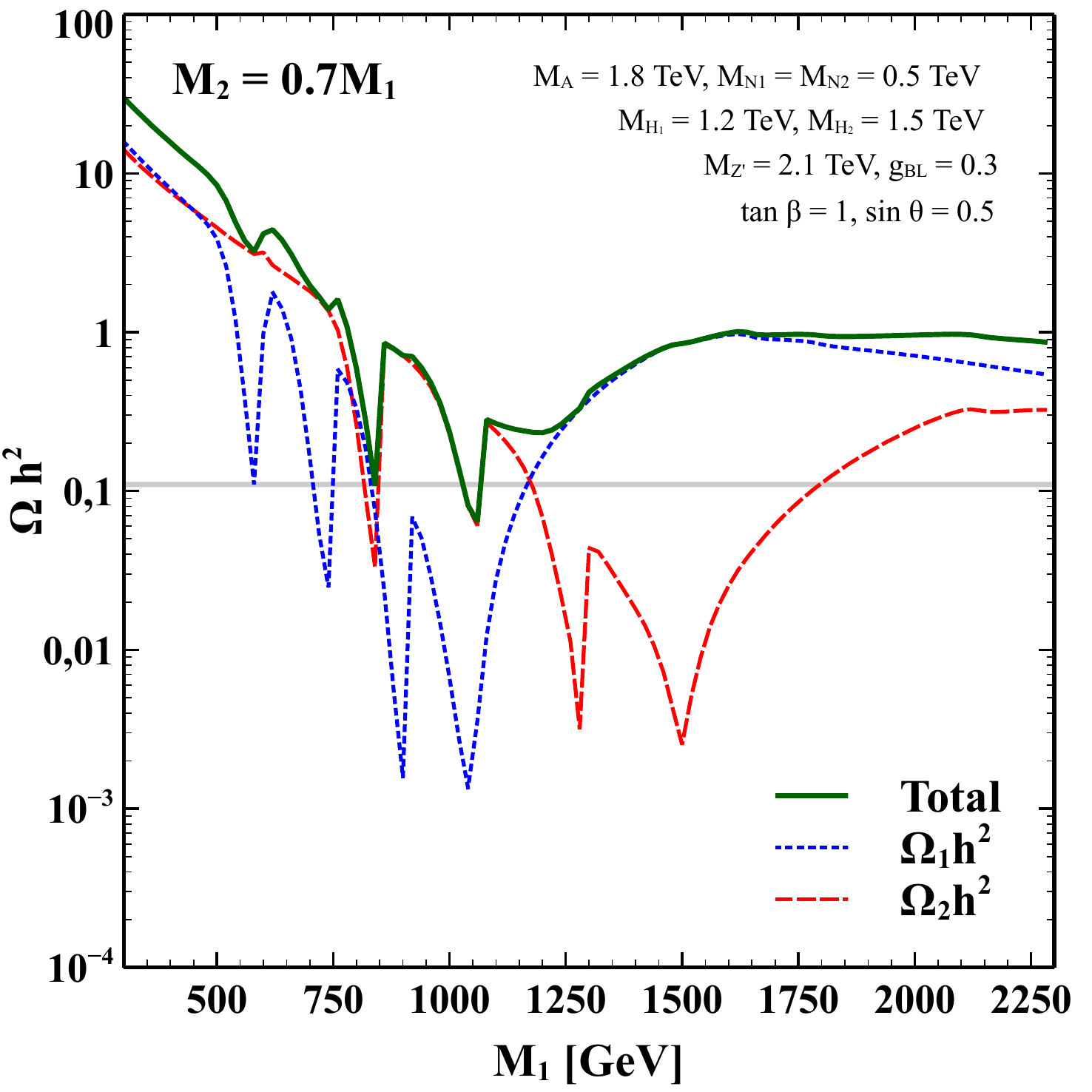}
\end{tabular}
\caption{\small The relic densities as a function of the $\psi_1$ mass for $M_2=1.3\,M_1$ (left), $M_2=M_1$ (center), and $M_2=0.7\,M_1$ (right). The other parameters were taken to be $M_{H_1}=1.2$ TeV, $M_{H_2}=1.5$ TeV, $M_A= 2.5$ TeV, $M_{N_1}=M_{N_2}=0.5$ TeV, $\mzprime=2.1$ TeV, $g_{BL}=0.3$, $\tan\beta =1$ and $\sin\theta = 0.5$.  The solid (green) line shows the total relic density, $(\Omega_1+\Omega_2)h^2$, while the dotted (blue) and dashed (red) lines show respectively $\Omega_1h^2$ and $\Omega_2h^2$. The horizontal band is the region consistent with Planck data.}
\label{fig:omega5}
\end{figure}

Let us now  investigate the dependence of the relic densities, $\Omega_1$ and $\Omega_2$, on the parameters of the model. The Yukawa couplings $y_{ij}$ play no role whatsoever in the DM phenomenology so, for the rest of the paper,  we will focus on the remaining $11$ parameters. In Fig.~\ref{fig:omega3} the relic densities are shown for a  case where both DM particles have the same mass, $M_1=M_2$. The rest of parameters were chosen as $M_{H_1}=M_{H_2}=M_A= 2.5$~TeV, $M_{N_1}=M_{N_2}=1$~TeV, $\mzprime=3.5$~TeV, $g_{BL}=0.5$, $\tan\beta =2$ and $\sin\theta = 0.5$. The dotted and dashed lines denote the relic densities of each DM candidate, $\Omega_1h^2$ (blue) and $\Omega_2h^2$ (red), whereas the solid (green) line  is their sum. For reference, the region compatible with Planck data~\cite{Aghanim:2018eyx} is shown as a gray horizontal band.  Several features are evident in this figure. The scalar and $Z'$ resonances, at a DM mass of $1.25$~TeV and $1.75$~TeV respectively, lead to a marked suppression of the relic density, as expected. At a DM mass of $2.5$~TeV, the annihilation into scalar final states opens up, giving rise to a reduction of the relic density. A smaller effect is also observed at a DM mass of $3$~TeV, where the annihilation into a $Z'$ plus a scalar becomes kinematically allowed. Notice that, over most of the mass range,  the relic density is dominated by $\psi_1$, indicating that the gauge interactions prevail. But as the DM mass increases, the Yukawa couplings $a$ and $b$ (associated with $\psi_1$ and $\psi_2$) become larger and the annihilations into scalars are enhanced --see Fig.~\ref{fig:feynman} (c). As a result, it is $\psi_2$ that contributes most to the relic density for masses above $2.5$~TeV. Also at the scalar resonance, $M_1\sim 1.25$~TeV, $\Omega_2$ turns out to be  larger than $\Omega_1$.

In Fig.~\ref{fig:omega1} the effect of varying the relation between the masses of the two DM particles  is illustrated. We now set $M_2=1.2\,M_1$ in the left panel and $M_2=0.8\,M_1$ in the right panel --the rest of parameters are the same as in Fig.~\ref{fig:omega3}. The resonances now occur at different values of the DM masses, yielding a more complicated function. We now observe, for instance, two regions where both particles have similar relic densities: 2~TeV $\lesssim M_1\lesssim 2.5$~TeV for the left panel and $M_1\lesssim 1$~TeV for the right panel.  As before, however, we notice  that the total relic density lies below the observed value only close to the resonance regions. 

The masses of the three scalars are expected  to be different in general, so there can be 4 different resonances for each DM particle. Fig.~\ref{fig:omega5} illustrates this possibility for different relations between the masses of the DM particles: $M_2=1.3\,M_1$ (left panel), $M_2=M_1$ (center panel), and $M_2=0.7\,M_1$ (right panel). For this figure, we chose a lighter spectrum, with the  $Z'$ at $2.1$~TeV, the two right-handed neutrinos at $0.5$~TeV, and the three scalars at $1.2$, $1.5$ and $1.8$~TeV respectively for $H_1$, $H_2$ and $A$. The rest of parameters were chosen to be $g_{BL}=0.3$, $\tan\beta =1$ and $\sin\theta = 0.5$. For $M_1=M_2$ (center panel), notice that the relic density is dominated by the $\psi_1$ practically over the entire mass range shown. In the left and right panel instead, there are  regions where it is $\psi_2$ that gives the dominant contribution to the relic density.

\begin{figure}[t]
\begin{tabular}{lr}
  \includegraphics[scale=0.5]{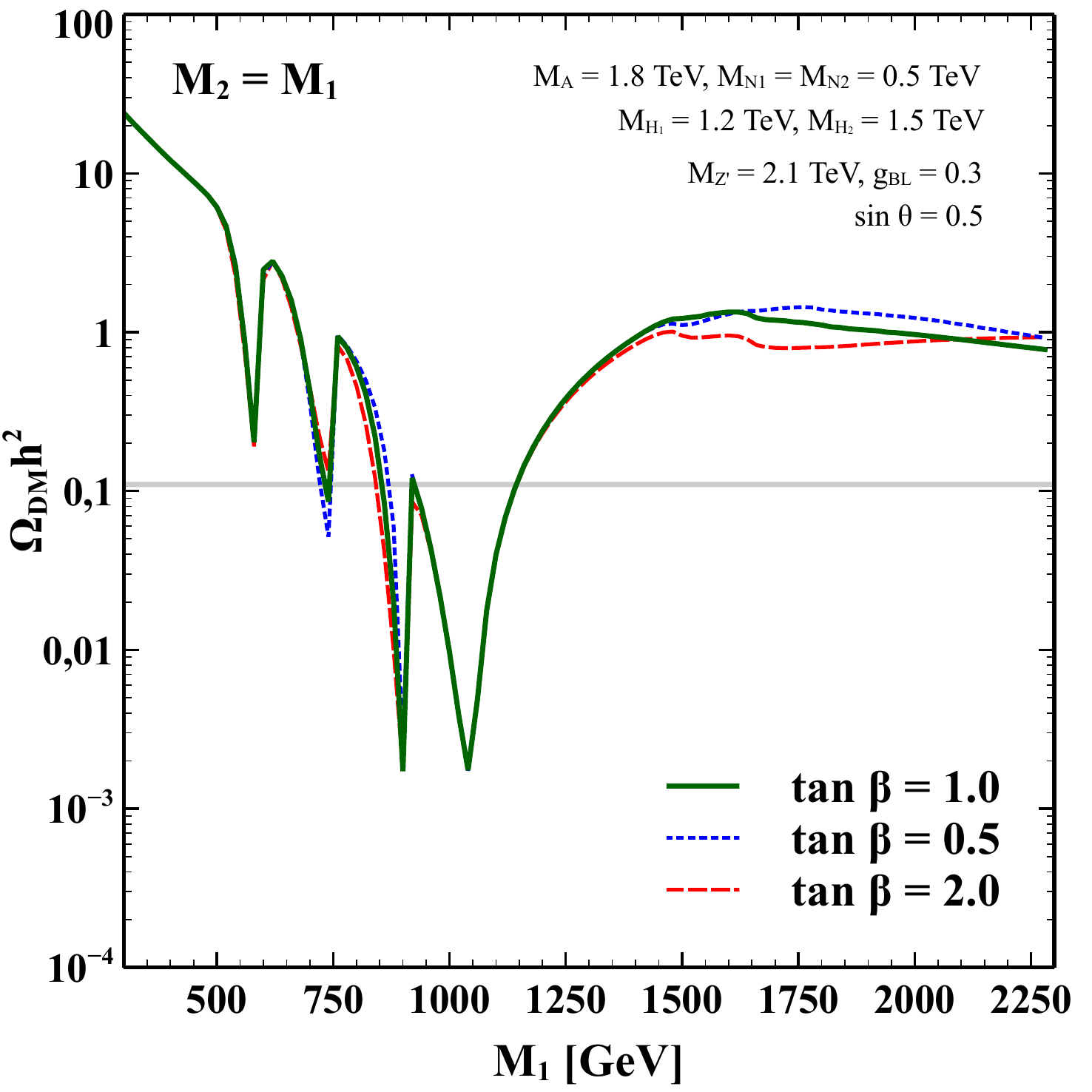} & \includegraphics[scale=0.5]{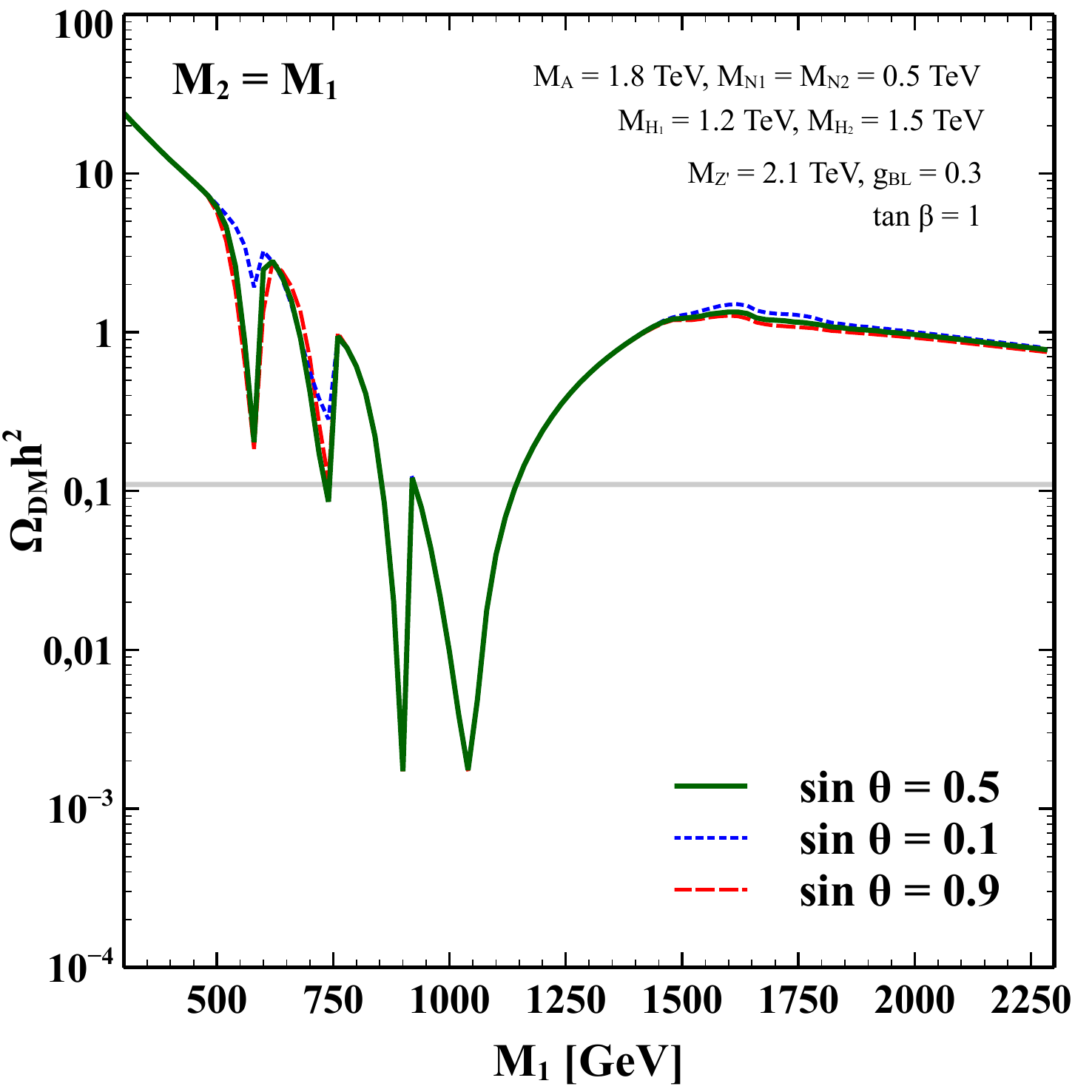}
\end{tabular}
\caption{\small The total relic density as a function of the $\psi_1$ mass for $M_2=M_1$  and different values of $\tan\beta$ (left) or $\sin\theta$ (right). The rest of parameters are the same as in Fig.~\ref{fig:omega5}.}
\label{fig:omegavars}
\end{figure}

$\tan\beta$ and $\sin\theta$ may also affect the relic density but only mildly and within specific regions of the parameter space, as depicted in Fig.~\ref{fig:omegavars}. The left panel shows the total relic density for three different values of $\tan\beta$: $1.0$ (solid line), $2.0$ (dashed line), and $0.5$ (dotted line). The remaining parameters are the same as in the previous figure. Notice that $\tan\beta$ modifies the relic density mostly at high DM masses, where the annihilation into scalars is important. The right panel displays the total relic density but now for three different values of $\sin\theta$: $0.5$ (solid line), $0.9$ (dashed line), $0.1$ (dotted line). From the figure we see that the  three lines mostly coincide, differing  slightly only at the $H_{1,\,2}$ resonances and at $M_1\sim 1.6$~TeV.  

The previous figures have illustrated the behavior of the relic densities with the different parameters of the model. What we would like to do now is to impose the relic density constraint and obtain the viable regions of this model. Then, we would like to examine whether such regions are also consistent with other current experiments, particularly the LHC and direct detection experiments, and whether they can be probed in the future. That is what we will do in the next section.

\section{The viable parameter space}
\label{sec:viable}

As we have seen in the previous section, the relic density can be obtained via gauge or scalar interactions, and agreement with the observed DM density  is achieved typically close to the resonance regions. To facilitate the exploration of the parameter space of this model and the determination of the regions consistent with the DM constraint, we will now limit our analysis to regions where the relic densities are obtained close to the $Z'$ resonance. Such parameter space points are also expected to be the most interesting ones, due to the correlations between the relic density, direct detection limits, and collider bounds on the $Z'$.

\begin{table}
\begin{tabular}{|c||c|}
\hline
Parameter & Range\\
\hline \hline
$M_{Z'}$ & $(0.3,22)$ TeV\\
$M_{1,\,2}$ & $(0.35, 0.65)~M_{Z'}$\\
$g_{BL}$ & $(0.001,1)$\\
$\sin\alpha$ & $(0.001,1)$\\
$\tan\beta$ & $(0.03,30)$\\
$M_{R1}$, $M_{R2}$ & $(0.2,10)$ TeV\\
$M_{H1}$, $M_{H2}$, $M_A$ & $(0.2,10)$ TeV\\
\hline
\end{tabular}
\caption{The parameters of our model and the ranges used in the random scan.}
\label{tab:scan}
\end{table}

We have done a random scan over the parameter space of this model, according to the ranges in Table \ref{tab:scan}. A point is considered viable if it is consistent with the LEP bound from Eq.~\eqref{eq:LEPII}, with the observed DM density, with perturbativity ($|\lambda|<1$, for all dimensionless scalar and Yukawa couplings), and with vacuum stability, Eq.~\eqref{eq:vacuum}. The  perturbativity and stability conditions were also analyzed at higher scales by using the two-loop Renormalization Group Equations of the model calculated using SARAH~\cite{Staub:2011dp, Staub:2013tta, Staub:2015kfa}. By following the criteria defined for example in Ref.~\cite{Escudero:2018fwn}, we evaluated the couplings at higher scales and  checked against Landau poles,  vacuum stability, and perturbativity (of dimensionless scalar and Yukawa couplings). If any of these conditions was broken at some scale $\Lambda$ below the highest mass in the model, such parameter space  point was discarded. Very few models in our sample (about 5\%) needed to be discarded due to this RGE criterion.

\begin{figure}[t]
\begin{center}
  \includegraphics[scale=0.75]{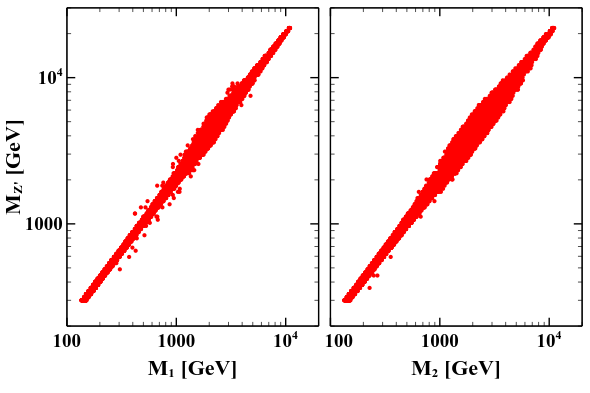} 
\caption{The viable points shown in the planes ($M_1$, $M_Z'$) and ($M_2$, $M_Z'$).
}
\label{fig:scanMDmMzB}
\end{center}
\end{figure}

\begin{figure}[t]
\begin{center}
  \includegraphics[scale=0.50]{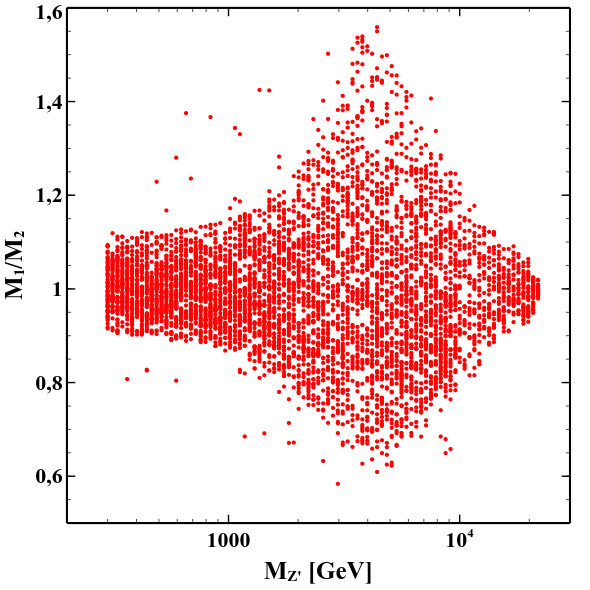}
\caption{The viable points shown in the plane ($M_Z'$, $M_1/M_2$).
}
\label{fig:scanMDmMz}
\end{center}
\end{figure}

In the following we will project the viable models into different planes so as to characterized them. To begin with, the correlation between the DM masses and the gauge boson mass is illustrated  in Figs.~\ref{fig:scanMDmMzB} and~\ref{fig:scanMDmMz}. We know that, by construction, $M_1$ and $M_2$ are necessarily close to $\frac12 M_{Z'}$ --see Table~\ref{tab:scan}-- so it is not surprising that all viable models lie  within a narrow band in this plane, as seen in Fig.~\ref{fig:scanMDmMzB} . Still, one can see that this band becomes slightly wider  in the central part, a feature that is more evident in Fig.~\ref{fig:scanMDmMz}, which shows a scatter plot of $M_1/M_2$ versus $M_{Z'}$. According to the ranges used in the scan, $M_1/M_2$ must lie between $0.5$ and $1.9$. From the figure, we see that for the viable models $M_1/M_2$ goes from  $(0.9,\,1.1)$ at low $M_{Z'}$ to $(0.6,\,1.6)$ for $M_{Z'}\sim 5$~TeV and then it narrows down again, reaching about $(0.95,\,1.05)$ for $M_{Z'}\sim 20$~TeV. Hence, the masses of the two DM particles tend to become identical at the upper end of $M_{Z'}$ (and of $M_{1,\,2}$).

\begin{figure}[t]
\begin{center}
  \includegraphics[scale=0.5]{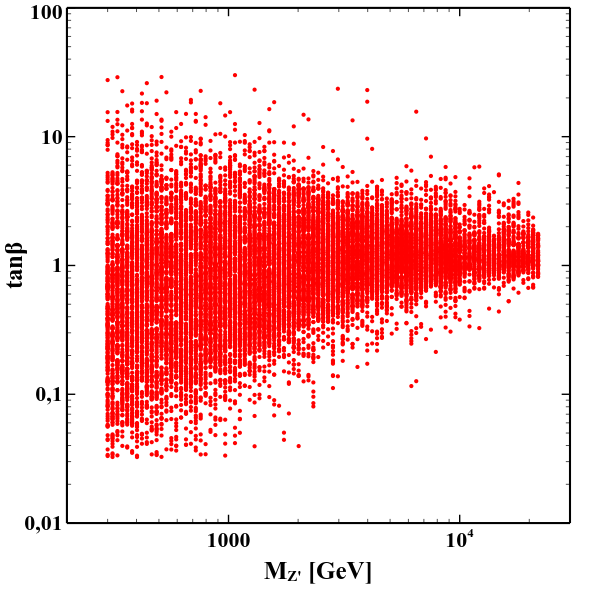}
\caption{The viable points shown in the plane ($M_Z'$, $\tan\beta$).
}
\label{fig:scanMzpvstb}
\end{center}
\end{figure}

The parameter $\tan\beta$ does not directly affect the DM relic density --see Eq.~\eqref{eq:cs}. It can modify  the viable parameter space, however, via the perturbativity and vacuum stability conditions. A scatter plot of $\tan\beta$ versus $M_{Z'}$ is shown in Fig.~\ref{fig:scanMzpvstb}. A priori, $\tan\beta$ can vary between $0.03$ and $30$. What we see from the figure is that this  range is actually realized only for a light $Z'$ (or equivalently for light DM particles), and that it becomes smaller as the masses increase. For $M_{Z'}$ around $20$~TeV, for example, $\tan\beta$ varies only between $0.5$ and $3$ approximately. That is, $\tan\beta$ tends to get close to $1$ at high masses.

\begin{figure}[t]
\begin{center}
  \includegraphics[scale=0.65]{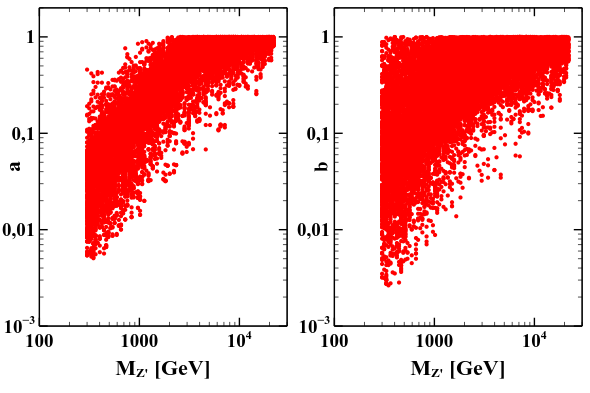}
\caption{The viable points projected onto the planes $(M_{Z'},\,a)$ and $(M_{Z'},\,b)$. $a$ and $b$ are the Yukawa couplings related to the DM particles --see Eqs.~\eqref{eq:TheModel} and~\eqref{eq:dmmasses}.
}
\label{fig:scandmcouplings}
\end{center}
\end{figure}

The Yukawa couplings, $a$ and $b$ in Eq.~\eqref{eq:TheModel}, associated with the DM particles are  displayed in Fig.~\ref{fig:scandmcouplings}. Since we imposed the perturbativity bound, all viable models feature $a,\,b<1$. From the figure we see that the minimum value of $a,\,b$ increases with $M_{Z'}$, which makes it more difficult to find viable models at high masses. Notice that there are plenty of models, over a wide range of masses, that saturate the perturbative bound we imposed.

\begin{figure}[t!]
\begin{center}
  \includegraphics[scale=0.5]{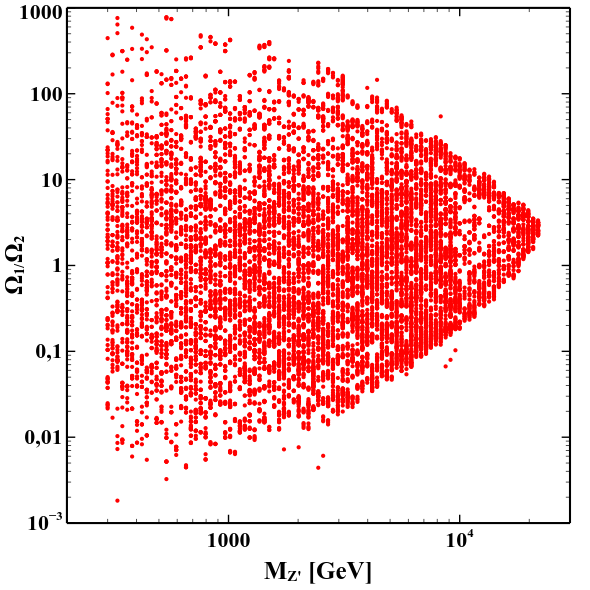}
\caption{The viable points projected onto the plane $\left(\mzprime,\,\frac{\Omega_1}{\Omega_2}\right)$.
}
\label{fig:scanomegas}
\end{center}
\end{figure}

Since we have two DM particles in this model, an important question to address is whether it is $\psi_1$ or $\psi_2$ that tends to dominate the DM density. Fig.~\ref{fig:scanomegas} displays the ratio $\Omega_1/\Omega_2$ for the set of viable models. One can see that the range of variation of $\Omega_1/\Omega_2$ gets reduced as $M_{Z'}$ increases, going from about $(10^{-2},\,10^3)$ at low $Z'$ masses to about $(1,\,4)$ for the highest masses found in our scan.  In particular, a scenario where both DM particles yield similar contributions to the observed DM density can be easily realized within this model.

This set of viable models we have found is further constrained by collider searches at the LHC and by DM detection experiments. Among the latter, it is the direct detection experiments that are expected to set the most relevant bounds in multi-component dark matter scenarios~\cite{Esch:2014jpa}, so we will focus on those.

\begin{figure}[t!]
\begin{center}
  \includegraphics[scale=0.5]{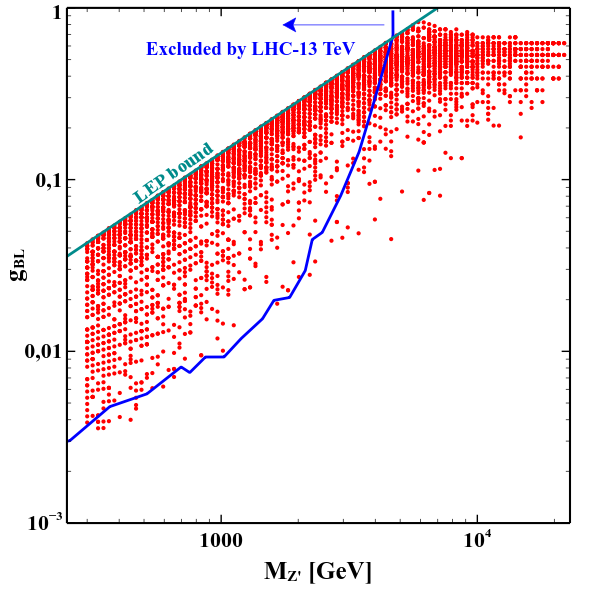}
\caption{The viable models projected onto the plane $(M_{Z'},\,g_{BL})$. For comparison, the region excluded by the LHC at 13~TeV \cite{Aaboud:2017buh,Escudero:2018fwn} is also displayed.
}
\label{fig:scanMdmgBL}
\end{center}
\end{figure}

Fig.~\ref{fig:scanMdmgBL} displays the viable points in the plane $(M_{Z'},\,g_{BL})$. First of all, notice that while  it is possible to find viable models over the entire range considered for $M_{Z'}$, there and no points with  $g_{BL}$ between $0.8$ and $1$. To give the correct relic density such points should feature $M_{Z'}\gtrsim 7$ TeV but it turns out that they are not consistent with the perturbativity bound we have imposed --the parameters $a$ and $b$ become greater than one. For $M_{Z'}\lesssim 6$~TeV,  the maximum value of $g_{BL}$ is set, at a given $M_Z'$, by the LEP bound from Eq.~\eqref{eq:LEPII}. The minimum value of $g_{BL}$ at a given $M_Z'$ is set instead by the relic density  constraint. The current bound from the ATLAS collaboration is shown as a solid (blue) line. It is based on dilepton searches with 36~fb$^{-1}$ of data at $\sqrt{s}=13$~TeV~\cite{Aaboud:2017buh,Escudero:2018fwn} (see Ref.~\cite{Sirunyan:2018exx} for similar CMS constraints on this channel). We see that this collider bound is quite strong, essentially excluding  $M_{Z'}$ below  $2$~TeV and partially constraining the region between $2$~TeV and $5$~TeV.  The region  $M_{Z'}\gtrsim 5$~TeV, instead, is not constrained by current LHC data.  Given that $M_{1,\,2}\sim M_{Z'}/2$~TeV (see Fig.~\ref{fig:scanMDmMzB}),  the region not currently constrained by the LHC corresponds to DM masses above $2.5$~TeV whereas DM masses below $1$~TeV are already excluded.

Additionally, in this model DM can scatter elastically off nuclei via a tree-level exchange of a $Z'$ in the $t$-channel. The spin-independent cross section for scattering of the DM off of a nucleon is given by
\begin{equation}
\sigma^\text{SI}_{1,\,2}\simeq \frac{m_n^2}{4\pi}\frac{g_{BL}^2\,g_{\psi_{1,\,2} V}^2}{M_{Z'}^4},
\end{equation}
where $m_n$ is the nucleon mass. Notice, in particular, that this cross section   is independent of the DM masses. According to this formula, the scattering cross section is  a factor $3.4$ larger for $\psi_2$ than for $\psi_1$ --see Eq.~\eqref{eq:dmcouplings}. It must be kept in mind, however, that, since we are dealing with a two-component DM scenario, the relevant experimental quantity is not simply the cross section but rather the product of the cross section times the DM fraction: $\Omega_1/\Omega_\text{DM}\,\sigma_{1}^\text{SI}$ and  $\Omega_2/\Omega_\text{DM}\,\sigma_{2}^\text{SI}$. At high masses, for instance, $\Omega_1$ tends to be slightly larger than $\Omega_2$ (see Fig.~\ref{fig:scanomegas}), which may compensate for the smaller value of the $\psi_1$ cross section.

\begin{figure}[t!]
\begin{tabular}{lr}
  \includegraphics[scale=0.5]{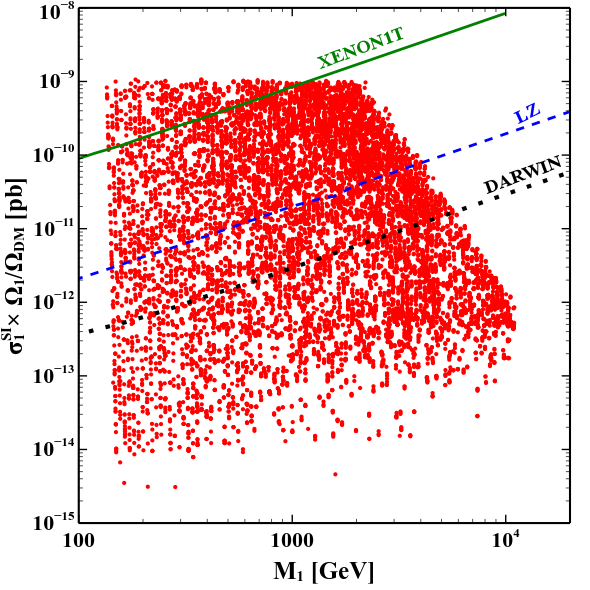} & \includegraphics[scale=0.5]{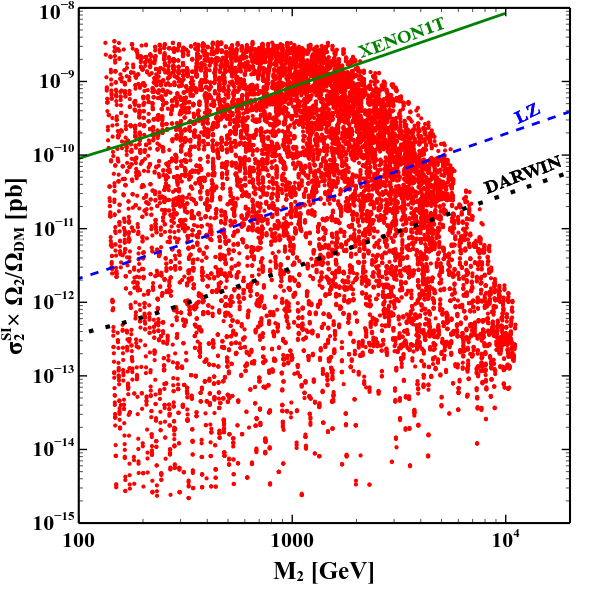}
\end{tabular}
\caption{The viable points projected onto the planes ($M_i$, $\sigma_i^\text{SI}\times\Omega_i/\Omega_\text{DM}$) for $i=1$ (left panel) and $i=2$ (right panel). The current limits from XENON1T are also shown, as well as the expected sensitivities of DARWIN and LZ.
}
\label{fig:scansip}
\end{figure}

Fig.~\ref{fig:scansip} shows the viable models in the planes $(\Omega_i/\Omega_\text{DM}\,\sigma_i^\text{SI},\, M_i)$ for both DM particles. As can be seen in the figure, $\Omega_i/\Omega_\text{DM}\,\sigma_i^\text{SI}$ varies between a maximum of $10^{-9}-10^{-8}$~pb  and a minimum of  $10^{-15}-10^{-14}$~pb. For comparison, we also show the current limit from XENON1T \cite{Aprile:2018dbl} as well as the projected sensitivities of LZ~\cite{Mount:2017qzi} and  DARWIN~\cite{Aalbers:2016jon}. 
A small region of the parameter space is already excluded by current direct detection experiments, and a much larger one lies within the expected sensitivity of future detectors. Future direct detection experiments, in particular, will  probe DM masses as high as $6$~TeV, well beyond the reach of current LHC searches ($\sim 2.5$~TeV). From the figure one can also see that  many viable models lie below the sensitivity of DARWIN  and will not be probed by future direct detection experiments.

\section{Conclusions}
\label{sec:conc}
The experimental evidence in favor of dark matter (DM) and neutrino masses compel us to look  for  physics beyond the Standard Model (SM). In this paper we proposed a new, and rather minimal, extension of the SM by an  $U(1)_{B-L}$ gauge symmetry.  This model has a very rich phenomenology: neutrino masses are generated via a TeV scale seesaw mechanism that leaves one  neutrino massless. Additionally, at colliders such as the LHC, it gives rise to interesting signals associated with the new gauge boson of $B-L$. Moreover, regarding DM, it automatically incorporates, without the need of any discrete symmetries, two DM particles, both of which are expected to contribute to the total DM density.   A novelty of  this model is that the anomalies are canceled  partially by  two right-handed neutrinos  and partially by the DM particles, providing a  connection between neutrinos and DM analogous to that one between leptons and quarks in the SM. The only other particles required in the model are two scalar fields that break the $B-L$ symmetry and give masses to the new fermions --of Majorana type for the neutrinos and of Dirac type for the two DM particles.  We described the model in detail and analyzed its most relevant phenomenological aspects.   The dependence of the relic densities with the parameters of the model was illustrated and the regions consistent with the DM constraint were identified. We showed that, after  imposing the current bounds from LHC and direct detection experiments,   the high mass region  of this model remains unconstrained.

\section*{Acknowledgments}
NB is partially supported by Spanish MINECO under Grant FPA2017-84543-P, and from Universidad Antonio Nariño grants 2017239 and 2018204. This project has also received funding from the European Union’s Horizon 2020 research and innovation programme under the Marie Skłodowska-Curie grant agreements 674896 and 690575, by Sostenibilidad-UdeA, and by COLCIENCIAS through the Grants 111565842691 and 111577657253.

\bibliographystyle{apsrev4-1}
\bibliography{newrefs}

\end{document}